\documentclass[traditabstract,twocolumn]{aa}
\usepackage{txfonts}
\usepackage{natbib}
\usepackage{lineno}
\usepackage{graphicx}
\usepackage{epsfig, longtable}
\bibpunct{(}{)}{;}{a}{}{,}
\newcommand{\fermi}{${\it Fermi~}$}

\begin{document}


\title{Implications for the structure of the relativistic jet from multiwavelength observations of NGC 6251}

\author{G.\ Migliori\inst{1,2} \and P.\ Grandi\inst{3} \and E.\ Torresi\inst{3,4} \and C.\  Dermer\inst{5}
\and J.\ Finke\inst{5} \and A.\ Celotti\inst{1} \and R.\  Mukherjee\inst{6} \and M.\ Errando\inst{6}
\and F.\ Gargano\inst{7} \and F.\ Giordano\inst{7,8} \and M.\ Giroletti\inst{9}}
\institute{SISSA/International School for Advance Studies, via
  Bonomea 265, I-34136 Trieste \and Harvard-Smithsonian Center for Astrophysics, 60
  Garden St, Cambridge, MA 02138, USA \email{migliori@head.cfa.harvard.edu}
  \and Istituto di Astrofisica e
  Fisica Cosmica - Bologna, INAF, via Gobetti 101, I-40129 Bologna
  \and Dipartimento di Astronomia, Universit\`a di Bologna, via
  Ranzani 1, I-40127 Bologna \and Space Science Division, Naval Research Laboratory,
  Washington, DC 20375, USA \and Department of Physics \& Astronomy, Barnard
  College, Columbia University, New York NY 10027, USA \and Istituto
  Nazionale di Fisica Nucleare, Sezione di Bari, I-70126 Bari \and
  Dipartimento di Fisica ``M. Merlin'' dell'Universit\`a e del
  Politecnico di Bari, I-70126 Bari \and INAF Istituto di
  Radioastronomia, via Gobetti 101, I-40129 Bologna}

\abstract{
NGC 6251 is a luminous radio galaxy $\approx 104$ Mpc away that was detected significantly
with the \emph{\it Fermi Gamma-ray Space Telescope}, and before that with EGRET (onboard the {\it Compton Gamma-ray Observatory}).
Different observational constraints favor a nuclear origin for the $\gamma$-ray emission.
Here we present a study of the spectral energy distribution (SED) of the core of NGC 6251, and give 
results of modeling in the one-zone synchrotron/SSC framework. The SSC model provides a good description of the radio to $\gamma$-ray emission but, 
 as for other misaligned sources, predicts a lower Lorentz factor ($\Gamma\sim2.4$) than typically found when
modeling blazars. If the blazar unification scenario
is correct, this seems to point to  the presence of at least two emitting regions in
these objects, one with a higher and one with a lower Lorentz factor.
The solution of a structured jet, with a fast moving spine surrounded by a slow layer, is explored and the consequences of the two models for the jet energetics and evolution are discussed.}

\titlerunning{Jet structure in NGC6251}
\authorrunning{Migliori et al.}
\maketitle


\section{Introduction }

The  \emph{Fermi Gamma-ray Space Telescope} has confirmed 
misaligned AGNs (MAGNs), including radio galaxies (RGs) and 
steep--spectrum radio quasars (SSRQs), as a new and important class of 
$\gamma$--ray emitters \citep{2010ApJ...715..429A,2010ApJ...720..912A}.
In the first year of activity, the Large Area Telescope \citep[LAT,][]{2009ApJ...697.1071A} onboard \fermi detected 11 MAGNs belonging to the 3CRR, 3CR, and  MS4 catalogs at 
178 and 408 MHz. At these low frequencies, the
Cambridge and Molonglo surveys 
favor the detection of radio galaxies over  blazars. 
Seven of the 11 \fermi MAGNs are nearby ($z\la 0.06$) Fanaroff-Riley type I \citep[FRI;][]{1974MNRAS.167P..31F} radio galaxies, while the rest are FRII radio galaxies.\\
\noindent 
The FRI radio galaxies detected with \fermi 
are significantly  less $\gamma$-ray
luminous than  BL Lac objects.
 The BL Lac objects have isotropic 100 MeV -- 10 GeV luminosities $L_\gamma \approx
10^{44} - 10^{46}$ erg s$^{-1}$, significantly higher than the FRI radio
galaxies, which have $L_\gamma \approx 10^{41} - 10^{44}$ erg
s$^{-1}$.
 They also have spectral slopes that are consistent with low- or intermediate-synchrotron-peaked 
blazars \citep[][]{2010ApJ...715..429A,2010ApJ...716...30A}.
This is consistent with AGN unified schemes, according to which an
increase in the inclination angle of the jet axis with respect to the
observer's line-of-sight implies a de-amplification of the observed
flux and a general shift in the emission to lower energies
\citep[][and references
  therein]{1984ApJ...280..569U,1995PASP..107..803U}.
The classification itself of misaligned sources is relative to our
  expectations of
  detecting in radio galaxies strongly beamed jet radiation from highly relativistic ($\Gamma\gtrsim10$) bulk flows,
  as in the case of blazars. 
Thus, for the sake of clarity, by misaligned here we do not mean that
the observer line-of-sight is outside the radiation cone of the source but rather
that it falls outside the (narrow) beaming cone
we would expect in the case of a highly relativistic core.\\
A full comparison with the expectations of the unification scenarios for AGNs
has not yet been considered, but the addition of a substantial number of misaligned
AGNs at 100 MeV -- 10 GeV energies now enables this to be made.  

\noindent
Here we present our study of the SED of the core of NGC 6251, identified
with 1FGLJ1635.4+8228 in the First \fermi-LAT Catalog 
\citep[][First \fermi-LAT Source Catalog, 1FGL, and
  First \fermi-LAT AGN Catalog, 1LAC, respectively]{2010ApJS..188..405A,2010ApJ...715..429A}. 
NGC 6251 is the fifth $\gamma$-ray brightest radio galaxy
in the MAGN sample with an integrated flux $F_{-9}=(36\pm8)$ in
units of $10^{-9}$ ph($> 100$ MeV) cm$^{-2}$ s$^{-1}$. 
It is associated with the EGRET source 3EG J1621+8203 \citep{2002ApJ...574..693M}, 
being after Centaurus A one of the most likely associations of an EGRET source with 
a radio galaxy (also of interest are NGC 1275, 3C 111, and M87). 
Being bright and relatively close \citep[$z=0.02471$][]{2003AJ....126.2268W},
NGC 6251 provides a nearby cosmic laboratory 
to explore the jet structure of radio sources hosting a supermassive
black hole.\\
After describing the source in more detail (Sect. 2), we describe the
multiwavelength datasets used to assemble the spectrum of NGC 6251 (Sect. 3).
In Sect. 4, we illustrate some uncertainties regarding jet orientation and location, before
presenting the results of one-zone synchrotron self-Compton (SSC) modeling in Sect. 5.
There we derive magnetic fields, outflow Lorentz
factors, and absolute power estimates for comparison
with blazars. Parameter comparisons between SSC models of  BL Lacs and 
FRI galaxies reveal the incongruity between these two classes of
sources, which we discuss in light of the unification scenario and explore
the feasibility of an alternative scenario.
In Sect. 6, we consider the connection between the jet power, the disk
luminosity, and the accretion.  
The results are summarized in Sect. 7.


\section{NGC~6251}\label{sect:the source}

 NGC~6251 is classified as a FRI radio galaxy
 \citep{1983MNRAS.204..151L} based on of its morphology and
 monochromatic radio luminosity\footnote{Throughout the paper, we adopt
   the cosmological parameter values $H_0=71$ km s$^{-1}$
   Mpc$^{-1}$, $\Omega_m = 0.27$, and $\Omega_\Lambda = 0.73$.} of 
$\approx 0.94\times 10^{31}$ erg s$^{-1}$ sr$^{-1}$ at 178 MHz
 \citep[][]{1977MNRAS.181..465W} \footnote{According to the
 classification proposed by \citet{1974MNRAS.167P..31F}, the flux density at the
   FRI/FRII boundary, corrected for the adopted cosmological values,
   is $1\times10^{32}$ erg s$^{-1}$ sr$^{-1}$.}.  Nevertheless, it
 exhibits an unusual radio morphology, with an $\approx 1\degr$ extension of bright radio
 emission to the northwest \citep{1977MNRAS.181..465W}, and
 characteristics typical of both
FRI and FRII radio galaxies \citep[see the radio maps at 327 and 608 MHz in][]{1997A&A...324..870M}.
 At a distance of  $\approx104$ Mpc, this corresponds to a linear extension of  $\approx 1.8$ Mpc  (1$\arcsec \cong 0.5$ kpc).\\ 
High resolution 1.48 and 4.9 GHz VLA studies 
reveal a radio core and a complex jet, that is bright and structured within $4.4\arcmin$ of
the core ($\sim113$ kpc), and then faint and curved at larger size scales \citep{1984ApJS...54..291P}. 
On the basis of radio images, \citet{2000AJ....120..697S} showed that the angle between the jet axis and the line of sight increases by $\ga 10\degr$, going from $33\degr$ at $50\arcsec$ from the core to $45\degr$ at $200\arcsec$.
The VLBI maps show an asymmetric core-jet radio structure aligned with the VLA jet \citep{1986ApJ...305..684J}. \\ 
Both non-thermal and thermal SED
components  contribute to the core 
emission of NGC 6251.  At wavelengths between $\approx 15$  and 30 $\mu$m, thermal dust emission appears
to dominate over an estimated $\sim$30\% nonthermal contribution to
the total mid-IR Spitzer flux \citep{2009ApJ...701..891L}.
Synchrotron radiation probably also accounts for the bulk of the optical to UV emission within $\approx 0.2\arcsec$ of NGC 6251's core \citep{2003ApJ...597..166C}.
The nuclear region in the optical band is complex. 
A warped dusty disk $\approx 1.43\arcsec$ in extension 
unevenly reflects UV radiation from the nucleus, as
seen in {\it Hubble Space Telescope} images \citep{1999ApJ...515..583F}. 
Ionized gas in the $0.3\arcsec$ region ($\approx 150$ pc) surrounding the nucleus
implies that the nucleus of NGC 6251 harbors a black hole with mass $\approx (4$ -- $8)\times 10^8$ M$_\odot$ \citep{1999ApJ...515..583F}.\\
NGC 6251 has been observed with space-based observatories at X-ray energies.
The unresolved core is the main X-ray emitter, 
although high-resolution imaging with
Chandra resolves  distinct X-ray emission in three different jet
regions \citep{2005MNRAS.359..363E}.  The detection of the Fe K$\alpha$ line is still debated \citep[][]{1997ApJS..113...23T,2003A&A...410..131G,2004A&A...413..139G,2005MNRAS.359..363E}.
An extended, $\sim100$ kpc X-ray  (k$_{\rm B}T \approx 1.7$ keV) thermal halo was detected with ROSAT \citep{1993ApJ...412..568B,1997A&A...324..870M}, Chandra \citep{2005MNRAS.359..363E}, and XMM-Newton \citep{2004A&A...414..885S}. A drop in the surface brightness of the X-ray halo, in positional agreement with the northern radio lobe, suggests that the lobe has evacuated a cavity in the surrounding X-ray gas \citep{2003NewAR..47..447K,2005MNRAS.359..363E}.\\ 
NGC 6251 has also been detected at hard ($>10$ keV) X-ray energies by Beppo-Sax \citep{2003A&A...410..131G,2006ApJ...642..113G} and INTEGRAL \citep{2005A&A...433..515F} and, as previously
noted, was proposed as a counterpart to the EGRET source 3EG J1621+8203 \citep{2002ApJ...574..693M}.  
Observation from \fermi confirmed that NGC 6251 is a GeV source \citep{2010ApJS..188..405A,2010ApJ...715..429A}.
The origin of the nuclear SED of NGC~6251 has been discussed in
several papers
\citep{2003ApJ...597..166C,2003A&A...410..131G,2005A&A...432..401G,2005A&A...433..515F},
which have concluded overall that the observed nuclear emission is likely to be dominated
by emission from a relativistic jet. 
The observed SED indeed shows the typical double-hump shape  characteristic of blazar sources.

\section{SED Data}

\subsection{Radio to optical/UV data}

The radio to UV data collected from literature and shown in Table
\ref{t1} are used to assemble the nuclear broadband SED of NGC~6251. 
We note that radio fluxes are taken at different angular resolution and sample nuclear regions of different dimensions. Moreover, the data are not simultaneous and variability could be an issue.  \citet{2005MNRAS.359..363E}, however, inspected VLBI core radio fluxes over 17 years and limited the maximum radio flux variability to a factor of $\leq 2$ \citep[see Table 9 in][]{2005MNRAS.359..363E}. 
Data in the microwave region are provided by the five-year Wilkinson
Microwave Anisotropy Probe (WMAP) catalog \citep{2009ApJS..180..283W} and by the published Planck Early Release Compact Source list  \citep{2011arXiv1101.2041P}.  
The IR flux (15-30 $\mu$m) is decomposed into $\sim$30\% synchrotron
jet emission extrapolated from the radio data and the remainder
thermal radiation
\citep{2009ApJ...701..891L}.\\ 
The non-thermal origin of the bulk of the nuclear optical emission is
  supported by the high degree of polarization of the UV emission
  (close to 50\%) and
  the position of NGC~6251 in diagnostic diagrams for radio galaxies, in
   common with the other FRI sources \citep[][and references therein]{2003ApJ...597..166C}.
The drop in the optical-UV band might be either real
  or cause by the dust extinction of the intrinsic optical and UV
  flux. It is important to establish wether the IR to X-ray flux can
  be ascribed to a single non-thermal process. The interpretation
  of the available data is partly controversial.
  \citet{2003ApJ...597..166C} argued against significant dust
  reddening because the progressive
  steepening of the optical-UV slope, related to the increase in the
  dust extinction with frequency, is not observed. By visual
  inspection, the optical nucleus appears to be very bright and unobscured by the dust
  lane. \citet{1999ApJ...515..583F} estimated a mean
  intrinsic visual extinction (A$_V$) of $0.61\pm0.12$ mag. The total
  extinction, which also accounts for the foreground reddening in the
  direction of NGC~6251, is A$_V=0.88\pm0.13$ mag. This
  value indeed relies on some assumptions, such as that
  the reddening law follows \citet{1989ApJ...345..245C} and $R_V\equiv
  A_V/E(B-V)$ is equal to 3.1. An upper limit of 1.0-1.5 to the value
  of A$_V$ was derived in \citet{2003ApJ...597..166C}.\\ 
In the SED (Figure \ref{f4}), we show both the optical fluxes for  no
absorption ({\it black solid triangles}) and after
de-reddening ({\it empty circles}) for A$_V=0.88$ mag \citep[using the extinction curves
  of][]{1989ApJ...345..245C}. The SED has a clear
double-peaked shape. However, even in the case of absorption ($A_V<1.01$), a
unique emission process for the IR to X-ray data seems unlikely as the
IR and optical fluxes do not lie on the extrapolation of the X-ray slope
to the lower frequencies \citep[see also Figure 13
  in][]{2005MNRAS.359..363E}. \\


\subsection{High energy  data}
 All data from XMM-Newton, Swift, and Chandra observations 
available in the public archives were re-analyzed  and 
included in the SED. 
For EGRET and  {\it Fermi}-LAT data, we use the spectra provided by
\citet{2002ApJ...574..693M} and  \citet{2010ApJ...720..912A}. 
\begin{table*}
\caption{X-ray data analysis results.}
\label{t2}
\begin{center}
\begin{tabular}{l cccccc}     
\hline
\hline                    
\noalign{\smallskip}
 Satellite                                                                & & XMM-Newton/PN                                                 &   Swift/XRT$ _1$                  &Swift/XRT$_2$                  &Swift/XRT$_3$    & XMM-Swift \\
Obs date                                                                & &2002-03-26                                                             &2007-04-06                           &2009-05-05                        &2009-06-05        &Combined fit$^{(a)}$  \\
\noalign{\smallskip}
\hline
\noalign{\smallskip}
k$_B$T (keV)                                                        & &0.6$\pm$0.2                                                           &--                                              &--                                          &--                                          & 0.8$^{+0.4}_{-0.2}$   \\
\\
norm$_{kT}^{(b,c)}$ ($\times$10$^{-5}$)                       & &3.0$\pm$2.0                                                          &--                                             &--                                           &--                                          & 4$\pm$2       \\ 
\\
N$_{H}$ ( $\times$10$^{21}$ cm$^{-2}$)       & &0.54$\pm$0.01                                                      &0.98$^{+0.08}_{-0.07}$      &1.4$^{+0.1}_{-0.1}$           & 0.6$^{+0.1}_{-0.6}$       &1.1$\pm$0.1\\
\\   
$\Gamma$                                                             & &1.89$\pm$0.04                                                     &2.00$^{+0.23}_{-0.21}$      & 2.20$^{+0.35}_{-0.31}$    &2.03$^{+0.35}_{-0.31}$   & 1.89$\pm$0.04\\
\\
norm$_{\Gamma}^{(c)}$ ( $\times$10$^{-3}$)                      & &1.18$\pm$0.05                                                     &0.85$^{+0.21}_{-0.16}$      &0.71$^{+0.27}_{-0.19}$     &0.60$^{+0.21}_{-0.15}$   & 1.18$\pm$0.05 \\                                                                              
\\        
n$_{XMM}^{(d)} $                                                          & &--                                                                               &--                                                &--                                                 &--                                            &1\\
\\
n$_{Swift1}^{(d)} $                                                        &  &--                                                                             &--                                                 &--                                                 &--                                         &   0.60$\pm0.04$\\
\\        
n$_{Swift2}^{(d)}$                                                         &  &--                                                                             &--                                                 &--                                                 &--                                             &   0.45$\pm$0.05\\
\\
n$_{Swift3}^{(d)}$                                                         &   &--                                                                            &--                                               &--                                                 &--                                                &  0.44$\pm$0.04\\
    
\\ 
$\chi^2$(d.o.f)                                                        &  & 410(372)                                                           &   21(29)                                & 9(11)                                       & 15(14)                                 & 455(430)\\
\\                                                       
\hline
\\
Flux$_{(0.5-2 keV)}^{(e)}$                                        & &2.7$\times$10$^{-12}$                                       &1.9$\times$10$^{-12}$      &1.6$\times$10$^{-12}$         &1.33$\times$10$^{-12}$&--                                                                       \\
\\
Flux$_{(2-10 keV)}^{(e)}$                                         & &3.6$\times$10$^{-12}$                                        &2.11$\times$10$^{-12}$  &1.42$\times$10$^{-12}$         &1.45$\times$10$^{-12}$ &--                                                                   \\
\noalign{\smallskip}
\hline
\end{tabular}
\tablefoot{
\tablefoottext{a}{Simultaneous fit of the four observations.}
\tablefoottext{b}{Normalization for APEC model is $norm=\frac{10^{-14}}{4\pi[D_A(1+z)]^2}\int n_en_HdV$ with the
    abundance table set to \citet{1989GeCoA..53..197A}.}
\tablefoottext{c}{In units of photons~cm$^{-2}$~s$^{-1}$.}
\tablefoottext{d}{Energy-independent multiplicative factor for
  the simultaneous fit.}
\tablefoottext{e}{Unabsorbed fluxes for the power-law component in
  units of erg~cm$^{-2}$~s$^{-1}$.}
}
\end{center}
\end{table*}

\subsubsection{X-ray spectral analysis}
The XMM--Newton observation of NGC~6251, performed in March 2002,  was analyzed using the SAS v.9.0 software and  available calibration files. 
We excluded time intervals affected by high background. After this
data cleaning, we obtained a net exposure and count rate of 8.7 ks and
1.867$\pm$0.015 count s$^{-1}$ for the pn, 13.9 ks and 0.553$\pm$0.006 count s$^{-1}$ for MOS1, and 13.5 ks and 0.556$\pm$0.006 count s$^{-1}$  for MOS2, respectively. The source and background spectra were extracted from circular regions of 27$^{\prime\prime}$ radius. The response matrices were created using the {\small SAS} commands {\small RMFGEN} and {\small ARFGEN}. 
The nuclear data are not piled up. Data were grouped into 25 counts
per bins in order to be able to apply the $\chi^{2}$ statistic. The
best-fit model for the pn (0.3-10 keV) data consists of an absorbed
power--law plus an {\small APEC} component which models a collisional gas emission ($\chi^{2}$=410 for 372 degrees of freedom). The parameter values, reported in Table \ref{t2}, agree with the results of \citet{2005MNRAS.359..363E}. The same results (not shown in Table \ref{t2}) were obtained using MOS data.  There is no significant evidence of the Fe K$\alpha$ emission line in the XMM data.\\
The Swift X--Ray Telescope observed NGC~6251 three times between April 2007 and May 2009.
The X-ray data were reduced using the on--line XRT data analysis  provided by the ASDC\footnote{http://swift.asdc.asi.it/}. Source spectra for each observation were extracted from a circular region of 20$\arcsec$ radius, while the background was taken from an annulus with an inner radius of 40$\arcsec$ and outer radius of 80$\arcsec$. The data were re-binned to 20 counts per bin in order to calculate $\chi^{2}$. All spectral fits were performed in the 0.5 -- 10 keV band. 
Data can be well described by an absorbed power-law with column density
slightly in excess of the Galactic value ($N_{H_{Gal}}=5.4\times10^{20}$
cm$^{-2}$). Unlike XMM Newton, the XRT spectra,  characterized by a
lower signal-to-noise ratio, do not require the addition of a soft
thermal emission component (see Table \ref{t2}).  We note that, during the 2002 observation, the source appeared in a higher state (see Sect. 3.2.2)\\
We analyzed only the most recent and longest Chandra observation of the source in November 2003
that was performed in ACIS--S (S3 and S4 chips) configuration. 
The Chandra pointing of September 2000 was not taken into account because 
in this observation the core fell in a chip gap.  
Data were reprocessed using the \emph{Chandra Interactive Analysis of Observation} ({\small CIAO} v.4.1) and the \emph{Chandra Calibration Database} {\small CALDB} 4.1. After correction for high particle background, the total exposure time is reduced to 45 ks.
The nuclear spectrum was extracted from a circular region of 5$\arcsec$ radius centered on the source. The background was chosen in an adjacent circular region with 12$\arcsec$ radius. Because of the brightness of NGC 6251 at X-ray energies,
the data are affected by pile-up, which was estimated to be $\approx 13$\%  using the {\small PIMMS} software\footnote{http://cxc.harvard.edu/toolkit/pimms.jsp}.
For this reason, we decided not to consider the Chandra nuclear fluxes in the SED.
As an extended extranuclear emission associated with the kpc jet is clearly visible in the Chandra image, 
a spectrum was also extracted for this component.
The accumulation  region is a box of 20$\arcsec$ and 5.3$\arcsec$ each side, far ($\approx 6\arcsec$) from the nucleus. 
A box of similar size was chosen to estimate of the background. A
good fit is obtained by an absorbed power law with
$\Gamma$=2.3$^{+0.26}_{-0.24}$ and N$_{H_{Gal}}=$5.7$\times$10$^{20}$
cm$^{-2}$. The photon index value is in good agreement with the one
($\Gamma$=2.29$^{+0.14}_{-0.13}$) reported in
\citet{2005MNRAS.359..363E} for the region containing the inner jet
(extending from about $1.5\arcsec$ to $27\arcsec$ from the
nucleus). The 0.5 -- 2 keV unabsorbed flux is 3.1$\times$10$^{-14}$
erg~cm$^{-2}$~s$^{-1}$, while the 2 -- 10 keV flux is
2.3$\times$10$^{-14}$ erg~cm$^{-2}$~s$^{-1}$. 
\begin{figure}
 \centering
  \includegraphics[angle=-90,scale=0.31]{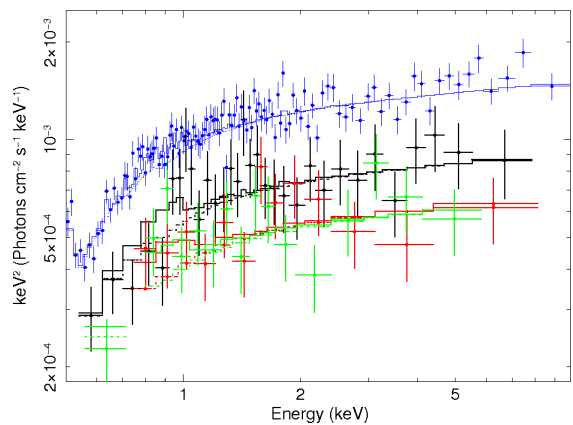}
   \caption{Unfolded spectral model of the combined
XMM-Newton and Swift datasets. Black points: Swift observation of April 2007;
red: Swift observation of May 2009; green: Swift observation of June
2009; blue: XMM-Newton observation of March 2002.}
  \label{f1}
\end{figure}
\subsubsection{High energy variability}
\begin{figure}
\centering
  \includegraphics[scale=0.35]{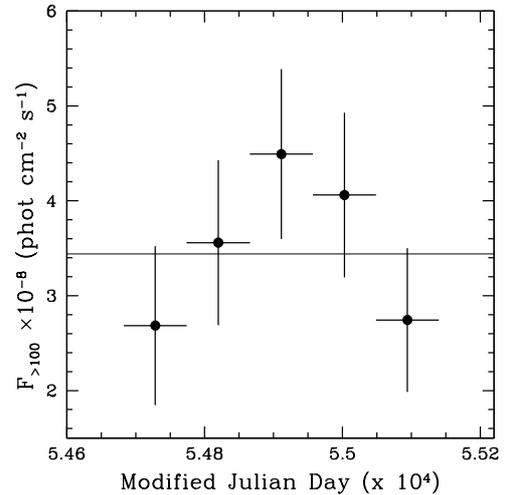}
   \caption{\fermi-LAT lightcurve of the source 1FGL1635.4+8228, associated with the radio
  galaxy NGC~6251, between 100 MeV and 100 GeV. The time is measured
  from 2008 August 4 and covers 15 months of \fermi operation. Each bin
  corresponds to 3 months of observations. The
  horizontal line is the constant flux.
}
  \label{f2}
\end{figure}

We first explored fast (hour-scale) temporal variability of the
  nuclear X-ray flux during the XMM and Swift observations producing a
  lightcurve in the 0.5 -- 10 keV band for each observation. None of
  the four lightcurves reveal a statistically significant change of flux. 
To investigate the presence of X-ray flux variability on
  timescales of months/years, we successively performed a simultaneous fit of the XMM and
  Swift spectra. 
The four datasets were simultaneously fit with a composite model, an
  absorbed power law, and a thermal ({\small APEC}) component,
 which correspond collectively to the best-fit model for the spectrum with the
  highest signal-to-noise ratio 
  (i.e. the XMM-Netwon dataset).  
An energy-independent multiplicative factor ($n$) allows us to compare
  the fits of the four datasets. The results of the simultaneous fit
  are reported in the last column of Table \ref{t2}. The values of
  the model parameters are driven by the XMM dataset. As also shown in
  Fig. \ref{f1}, there is an evident flux variation over the entire 0.5 -- 10 keV band between
  the XMM and Swift observations, expressed by the different $n$
  values. 
  The X-ray flux decreases by about $40\%$ between the XMM and the
  first Swift observations and then an additional $\sim15\%$ by
  the two last Swift observations (the values of $n$ parameter for
  the three observations being $n_{XMM}=1$, $n_{Swift1}=0.60\pm0.04$
  and $n_{Swift2}=0.45\pm0.05$ respectively).  
The close to two years between the first and the second Swift
  observations, performed on April 6, 2007 and May 5, 2009,
  respectively, provide us with the shortest interval during which X-ray flux
  variability has been detected. Thus, we consider this as an
  upper limit to the  X-ray flux variability timescale.\\  
No time variability in the $\gamma$-ray band was detected 
in the  first 15 months of LAT observations of the \fermi source
1FGLJ1635.4+8228, which is associated with NGC~6251 
\citep{2010ApJ...720..912A}. In Fig. \ref{f2}, we show  the 100
  MeV -- 100 GeV lightcurve of the 15-month LAT dataset generated by dividing
  the total observation period in 5 time intervals of 3-month
  duration. \\
  We considered
  longer timescales when comparing the LAT and EGRET results.
The LAT flux  $F_{-9}= 36\pm 8$ is lower in comparison with the
$\gamma$-ray flux measured by EGRET, $F_{-9}= 74\pm 23$. 
However, the large uncertainties in the fluxes make it difficult to firmly establish whether this is related to an actual variation in the $\gamma$-ray flux of the source on timescales of years rather than to contamination of the EGRET flux from other sources.
At a distance in the range of $\approx2\degr$--$5\degr$ from
1FGLJ1635.4+8228, there are at least three other $\gamma$-ray sources
in the 1FGL  with a flux similar to or brighter than 1FGLJ1635.4+8228, none of which has an EGRET counterpart.

\section{Observational constraints}

We must now consider our observational constraints when developing a multiwavelength 
database for modeling, including the relative imaging capabilities of different 
detectors,   epochs of different observations, separate spectral components, 
and questions about whether there is either intrinsic or Galactic extinction.

\subsection{Location of the $\gamma$-ray emission region}
The construction and interpretation of NGC 6251's SED depends on the site of production 
of the $\gamma$-ray photons. Because of the limited imaging capability
of the {\it Fermi}-LAT, 
with a PSF of $\approx 0.6\degr$ at 1 GeV (Atwood et al. 2009), \fermi can distinguish
the central regions from sites coincident with the extended radio
structures in NGC 6251, but is unable to resolve details on length-scales $\la 3\arcmin$ ($1\arcmin$ corresponds
to $\approx 30$ kpc). 
Radio galaxies are expected to be sources of $\gamma$-ray emission
produced in both the compact core and
inner jet \citep{2008Natur.452..966M,2010Natur.463..919F},
as well as emitting extended $\gamma$-ray emission from, at least, photons of the cosmic microwave and
extragalactic background light (CMB and EBL respectively) that are Compton-scattered by the
radio-emitting electrons in the lobes \citep{2007AIPC..921..325C,2008ApJ...686L...5G,2009MNRAS.393.1041H}. \\
If the source varies, simultaneous multi-frequency VLBI,
X-ray/Swift, and \fermi campaigns
\citep{2010arXiv1006.3084S,2010arXiv1006.3243G} can help to constrain the origin of 
the $\gamma$-ray emission as well as the physical parameters of the
region. For blazars, results based on high frequency radio
observations, rapid $\gamma$-ray variability, and correlated variability in different wavebands make locations in the vicinity of 
the bright radio core the favored candidate for the site of intense $\gamma$-ray production.\\
The same approach might be more difficult for MAGNs. Seven of the 11 MAGNs have a flux of $F_{-9}< 40$, and 
two of them (3C78/NGC1218 and PKS 0625-354) have $F_{-9}\la 5$. 
With the exception of NGC~1275, no significant $\gamma$-ray
variability has been detected so far
with the {\it Fermi}-LAT  in MAGNs
\citep[][]{2009ApJ...699...31A,2010ApJ...715..554K,2010ApJ...715..429A},
but, at these low $\gamma$-ray fluxes, we are unable to detect variability even when present. \\
Significant contributions from other components can further dilute and hide the flux variability.
For the first time, {\it Fermi}-LAT has detected $\gamma$-ray emission
from the lobes of the radio galaxy Cen A \citep{2010Sci...328..725F}:
the $>100$ MeV emission from the lobes is a large fraction
($\ga 1/2$) of the total flux, with the fluxes of the northern lobe, the southern lobe and the core, being $\approx 0.77\pm0.3\times 10^{-7}$, $1.09\pm0.3 \times 10^{-7}$, and $1.50\pm0.3 \times 10^{-7}$ ph  cm$^{-2}$ s$^{-1}$ respectively.\\ 
As discussed in Sect. 3.2.2, $\gamma$-ray variability on month timescales
has not been observed by \fermi, although flux variability over several
years is not completely ruled out.
Since no variability has yey been clearly detected, a significant fraction of
the $\gamma$-ray flux could originate from the lobes.
The $\gamma$-ray spectral index is soft
($\Gamma_{\gamma}=2.52\pm0.12$), implying that most of the power
detected in the {\it Fermi} band is dominated by photons in the 100 MeV -- 10 GeV range. At these energies, the LAT PSF is unable to resolve
  the core from the lobes.
As a rough
estimate, we can rescale the $\gamma$-ray flux of Cen A lobes at the proper luminosity
distance ($\sim 104$ Mpc) and lobe dimensions of NGC~6251. 
We use the $\gamma$-ray brightest lobe of Cen A (i.e. the southern
lobe) and assume a simple spherical geometry for the NGC~6251 lobe
with a diameter $D=1.9\ {\rm Mpc}/2=850$ kpc, equal to half of the
NGC~6251 linear dimension, and thus an overestimate of the actual lobe dimensions. In this way, the $\gamma$-ray flux
rescaled to the NGC~6251 lobe is $F_{-9}\approx8$,  comparable to the statistical error in the
LAT flux of the source associated with NGC~6251 ($F_{-9}=36\pm8$). 
There are indeed uncertainties in this estimate: we assume a
similar SED for the lobes of the two radio galaxies while this cannot
be assessed not even in the radio-IR band as the lobes of NGC~6251 are
not well sampled.
For this reason, we discuss the effects on the nuclear SED
modeling of a possible extended
contribution in Sect. 5.\\
Even though $\gamma$-ray emission on the hundred parsec scale might
also contribute, the non-thermal nuclear emission of the core likely
dominates the source SED at radio, X-ray, and $\gamma$-ray wavelengths. In Fig.\ 12 of \citet{1986ApJ...305..684J}, the core and jet-knot radio fluxes are compared at different frequencies: the core is optically thick below $\sim 13.5$ GHz, and becomes dominant at higher radio frequencies.
The X-ray spectral analysis, using high resolution Chandra data, has
shown that the bulk of the X-ray emission comes from the nuclear
region ($r\leq 4\arcsec$), while the knot accounts for
$\la 10$\% of the total observed X-ray flux
\citep[][]{2005MNRAS.359..363E}.

\subsection{Jet orientation}

\begin{figure}
 \centering
  \includegraphics[scale=0.4]{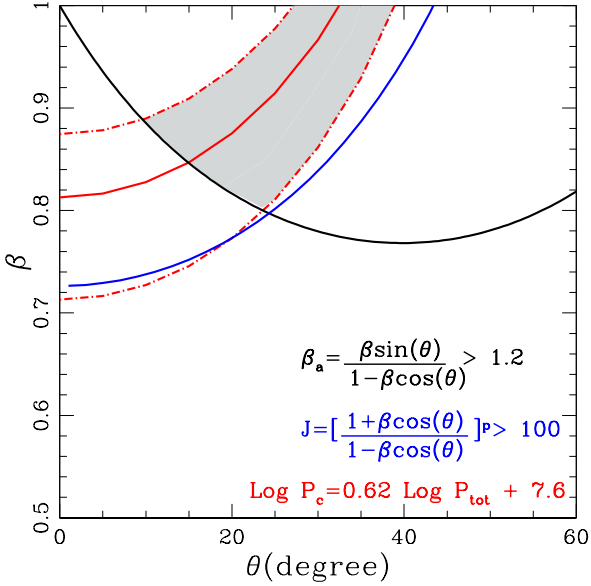}
\caption{Constraints on the angle $\theta$ between the jet and the line of sight and the jet velocity $\beta$ in units of  the speed of light.
Blue and black curves correspond to  $J=100$ and $\beta_a=1.2$, respectively.
The red curve is from the $P_c$ versus $P_{tot}$ relation. The dot-dashed red lines are obtained allowing the core density to vary by a factor two.
The gray area corresponds to the permitted values of $\beta$ and $\theta$. }
  \label{f3}
\end{figure}

Another  key question for MAGNs  is the jet orientation, which
is expected to be a large angle in objects with 
no extreme blazar characteristics.  
The jet inclination angle, a difficult quantity to observationally establish, 
must be measured before constraining the physical properties of
relativistic sources.
Here we consider a range of jet inclinations for NGC 6251
based on different observational quantities: the jet sidedness ($J$), 
the VLBI apparent velocity $\beta_a=v_a c$ of the jet knots,
and the core dominance.
If an intrinsic symmetry is assumed, the jet and counter-jet brightness ratio J can be expressed in terms of the jet velocity  ($\beta=v c$) and orientation ($\theta$):  $J=[(1+\beta \cos\theta)/(1-\beta \cos\theta)]^{(2+\alpha)}$, where $\alpha$ is the radio spectral index, defined such that $S(\nu)\propto \nu^{-\alpha}$, and assumed here to be $\alpha=0.5$. In a similar way the apparent transverse velocity of  a relativistic moving blob $\beta_a$ is related to $\beta$ and $\theta$ via $\beta_a= \beta \sin \theta/(1-\beta \cos \theta)$ \citep{1995PASP..107..803U}. 
\citet{2001ApJ...552..508G} found a general correlation, albeit with significant spread, between the core and total radio power in radio galaxies $\log P_c = 0.62 \log P_{tot}$ + 7.6,  where 
$P_c$   is the arcsecond core radio power at 5 GHz and $P_{tot}$ 
is the total radio power at 408 MHz in units of W Hz$^{-1}$. 
This relation can be used to derive upper and lower limits to  $\beta$ and $\theta$ \citep[for details, see][]{1994ApJ...435..116G},  allowing the core density to vary within a factor of 2.
Thus,  if  $P_c$ and $P_{tot}$ are known, and values or upper limits
to $J$ and $\beta_a$   are  available,  a narrow region of permitted
values can be defined  in the  $\beta$ versus $\theta$ plot. This
implicitly assumes that  the pattern speed inferred from the apparent motion and the bulk motion of the emitting plasma do not differ  \citep[][and references therein]{1995PASP..107..803U}.\\
In NGC 6251, the counter jet observed in VLA maps \citep{1984ApJS...54..291P} disappears on mas scales.  
\citet{2002ApJ...580..114J} investigated the possibility that free-free absorption by an ionized accretion disk 
could hide the receding jet, but concluded on the basis of the high electron density required that relativistic boosting effects better explain the observations.  In line with their considerations,  we assume that the brightness ratio of $\approx$ 100:1 measured within 6 mas of  the core 
is a plausible lower limit to $J$ \citep{1994ApJ...427..221J}.
The superluminal velocity of the jet has not been measured in NGC 6251,
only a lower limit of $\beta_a>1.2$ having
been reported \citep{1994ApJ...427..221J}.  
The values of $P_c$ and $P_{tot}$ that we considered here  to estimate  the jet
inclinations and velocities allowed by the core dominance correlation,
were calculated from the 5 GHz core flux $F_c(5\ {\rm GHz})$=0.4 Jy
\citep{1984ApJS...54..291P} and the total 408 MHz flux $F_{tot}(408\ {\rm MHz})$=5.3 Jy \citep{1977MNRAS.181..465W}.\\
The final $\beta$-$\theta$ plot  is shown in Fig. \ref{f3}. The permitted values, shown by the gray region, 
are within the area delimited by $J>100$, $\beta_a>1.2 $, and the core dominance relation.
Only  bulk motions larger than $\sim 0.78 $ and  inclination angles
in the range of about $\approx 10\degr$ -- $40\degr$ are compatible with the
observed properties of the NGC~6251  jet.

\section {SED modeling}

\subsection{One-zone SSC model}

\begin{figure*}
 \centering
  \includegraphics[scale=0.4]{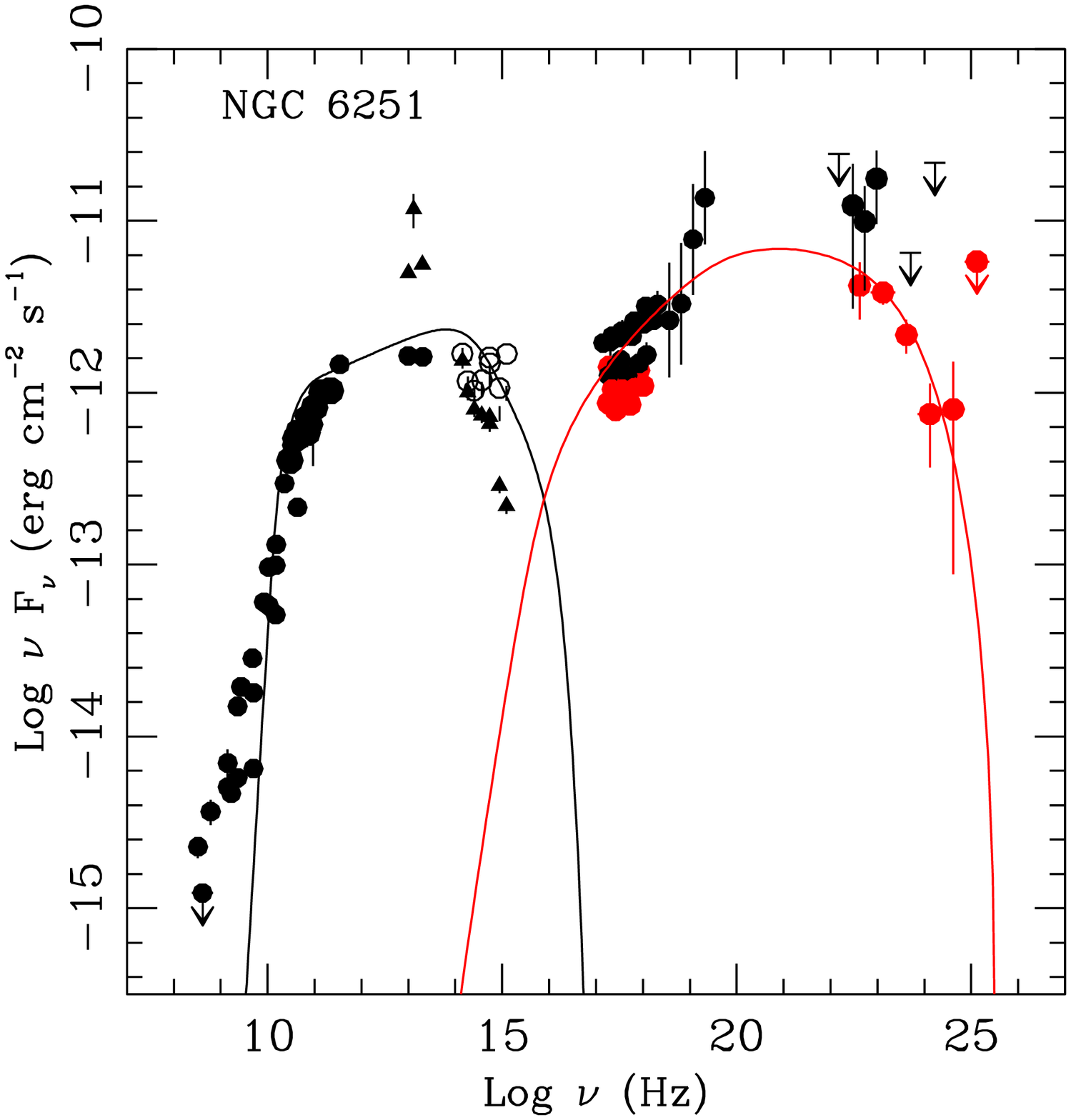}
   \includegraphics[scale=0.4]{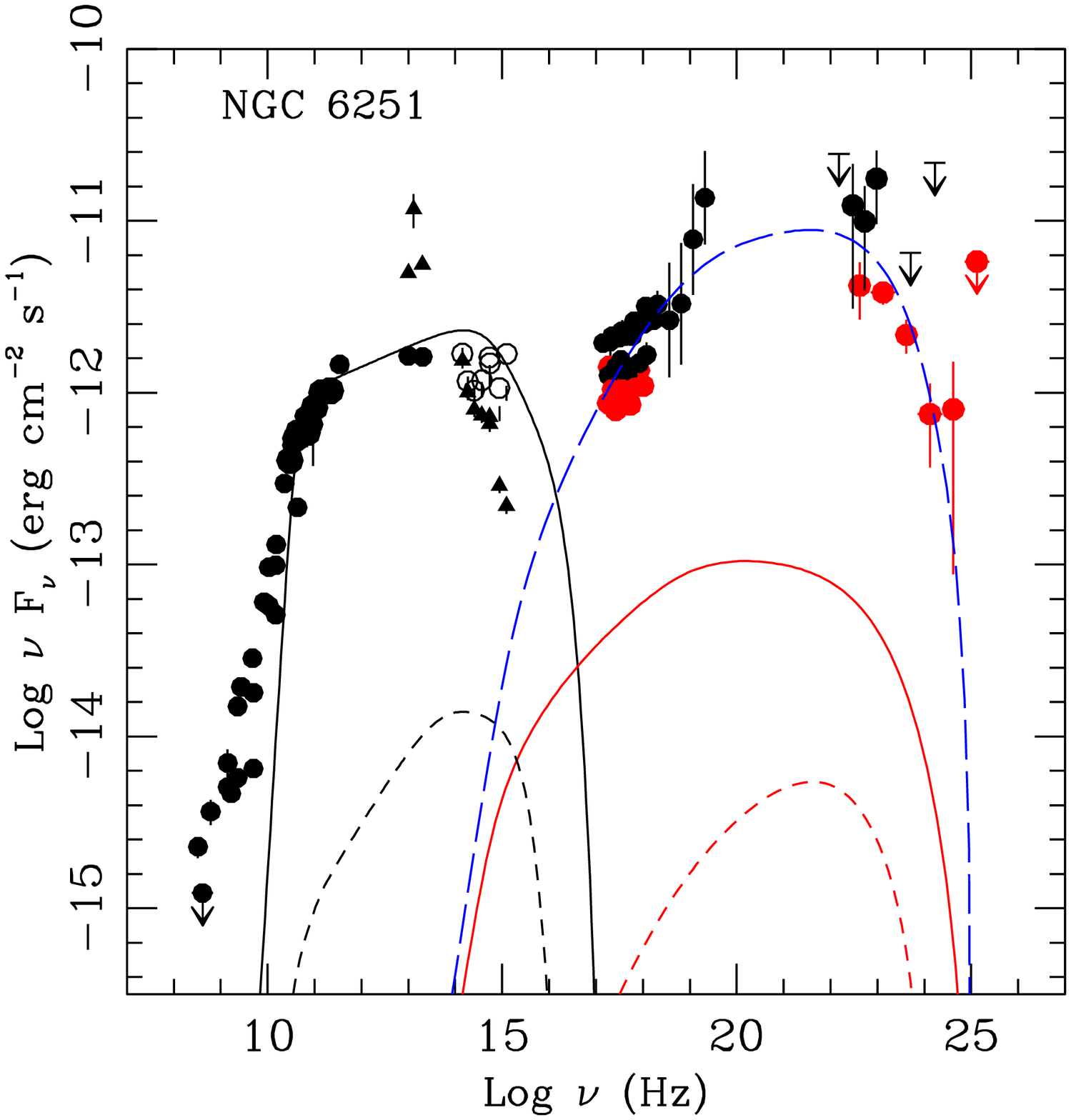}
\caption{Nuclear broadband SED of NGC~6251 compiled using multi-epoch
  data. Radio to optical-UV data are reported in Table \ref{t1}. Solid
  black triangles are the total IR flux. The solid IR points are the
  extrapolated non-thermal synchrotron fluxes given in
  \citet{2009ApJ...701..891L}. Empty optical-UV data are dereddened
  (see text). X-ray emission is presented in Sect. 3.2: black dots are
  for XMM-Netwon data and red from Swift satellite. Black $\gamma$-ray
  points correspond to EGRET flux \citep{2002ApJ...574..693M}.  Red
  $\gamma$-ray points are \fermi data
  \citep{2010ApJ...715..429A}. {\it Left panel:} the SED is modeled
  with a one-zone SSC model (solid black line: synchrotron curve,
  solid red line: IC emission). {\it Right panel:} the SED is modeled
  with the spine-layer model illustrated in
  \citet{2005A&A...432..401G}. Solid black and red lines reproduce the
  SSC emission of the layer. Dashed black and red curves are the SSC
  model for the spine. The long-dashed blue line is the IC emission of
  the spine synchrotron photons scattered by the layer relativistic electrons.}
  \label{f4}
\end{figure*}

In the context of the emission from a relativistic jet, 
the double-hump shape of the nuclear SED is accounted for the  
nonthermal  synchrotron and inverse Compton (IC) radiation. 
The radio to optical/UV radiation originates from the synchrotron mechanism 
by relativistic electrons, while the same electrons are supposed to 
Compton-upscatter seed-photons to X-ray and $\gamma$-ray energies.\\
\noindent
We first model the SED of NGC 6251 with a one-zone synchrotron self-Compton (SSC) model, where the 
seed photons for Compton scattering are the local synchrotron photons themselves. 
We assume a homogeneous spherical blob.
VLBI observations \citep{1986ApJ...305..684J} place an upper limit of $\la 0.20$ mas ($\approx 3\times 10^{17}$ cm) on  the size of the nuclear region. 
There is no clear evidence of flux variability on short timescales that could provide more stringent limits.  
VLBI radio fluxes vary of close to two on timescales of years \citep{2005MNRAS.359..363E},
corresponding to a region radius $R\sim 10^{18}$ cm $\sim 1$ pc. 
\citet{2004A&A...413..139G} reported the detection of short-time ($\sim 10^4$ s), low-amplitude X-ray flux variations. \citet{2005MNRAS.359..363E}, analyzing XMM-Newton and Chandra data, concluded that variations in 2 -- 10 keV fluxes are plausible but uncertain. \\
We derive upper limits similar to the radio ones on the basis of our X-ray temporal
analysis of XMM and Swift observations (Sect, 3.2.2). The assumption is that the nuclear X-ray emission is produced in the
same region as the radio and $\gamma$-ray radiation.\\
\noindent
As discussed in Sect. 4.2, the
jet/counter-jet flux ratio, apparent velocity, and the core-dominance relation 
define a range of allowed values for the core jet speed $\beta $ and relative inclination $\theta$.
We consider as reference values for the modeling $\theta=25\degr$
and $\beta=0.91$, values that fall roughly at the center of the area
of permitted values (see Fig. \ref{f3}). 
The resulting Doppler  factor is $\delta=2.4$. This value
agrees with the constraints inferred by assuming that
the region is transparent to $\gamma$-ray absorption via the $\gamma\gamma$ to e$^+$e$^-$
process, $\delta\geq0.44$ \citep[calculated using Equation A4
  in][Appendix]{2010ApJ...719.1433A}.\\
Looking at the broad-band SED (Fig. \ref{f4}), the synchrotron and Compton peaks can be approximately located in the frequency intervals $10^{13}$--$10^{14}$ Hz and $10^{21}$--$10^{22}$ Hz, respectively. 
\citet{2003AJ....126.2677Q} report a spectral index $\alpha_{O-S}=1.1$ between the sub-millimeter (870 $\mu$m) and the optical band, while the IR-to-UV slope between 16,000 and 2,200 \AA\ is $\alpha_{IR-UV}=1.75\pm0.16$ \citep{2003ApJ...597..166C}. 
The spectral index values in the high energy band,
$\alpha_X=0.88\pm0.04$ in the range 2-10 keV and
$\alpha_{\Gamma}=1.52\pm0.12$ in the {\it Fermi}-LAT frequency
interval, are slightly flatter than, but still consistent with, the
sub-mm/UV $\alpha$ values. \citet{2003AJ....126.2677Q} noted that dust extinction can partially reduce the flux of the optical continuum in the core, and determined the spectral steepening.\\ 
The four spectral indices give the slope below ($\alpha_{O-S}$,
$\alpha_X$) and above ($\alpha_{IR-UV}$, $\alpha_{\Gamma}$) the
synchrotron and Compton peaks, respectively.
Therefore, as a reasonable approximation, we assume the same spectral indices for the synchrotron and IC curves below ($\alpha_1=0.88$) and above ($\alpha_2=1.52$) the peaks, and adopt a broken power-law to describe the electron energy distribution (EED)
\begin{equation}
N(\gamma)=\left\{
\begin{array}{lr}
K\gamma^{-p_1} &  for\ \gamma_{min}\leq\gamma<\gamma_b,\\
K\gamma_b^{p_2-p_1}\gamma^{-p_2} & for \ \gamma_b\leq\gamma\leq\gamma_{max},
\end{array} \right.
\end{equation}
where $\gamma$ is the electron Lorentz factor in the range
$\gamma_{min}\leq \gamma \leq \gamma_{max}$, $K$ is the EED
normalization, $p_1=2\alpha_1+1$ and $p_2=2\alpha_2+1$ are the low and
high energy EED spectral indices, and $\gamma_b$ is the energy break of the distribution. From the ratio of the synchrotron ($\nu_s$) and IC ($\nu_c$) peak frequencies
\begin{equation}
\gamma_b=\Bigg(\frac{3\nu_c}{4\nu_s}\Bigg)^{1/2}\;,
\end{equation}
which yields values for $\gamma_b$ spanning the range between $\approx 3\times10^3$ and $\approx 3\times10^4$.\\
The best SSC model `fit' for the nuclear SED of NGC~6251 is shown in Fig. \ref{f4} ({\it left panel}), 
and the corresponding model parameters (in the comoving frame) are reported in Table \ref{t3}.\\
Interestingly the nuclear SED can be broadly reproduced by assuming a mildly relativistic
motion ($\Gamma=2.4$) , i.e. significantly lower than that required to account for
several observational properties of  blazars (i.e. the apparent superluminal
  motion observed on milliarcsecond scales, see
  \citealp{2004ApJ...609..539K,2009AJ....138.1874L}, and SED modeling,
  \citealt{2010MNRAS.401.1570T}), but with values of the source size ($R=1.2\times10^{17}$ cm) and magnetic field ($B=0.037$ G) consistent with
those inferred for BL Lac objects.
Similar results have also been found when a single-zone
synchrotron/SSC model is applied to the nuclear SED of other members
of the MAGN sample detected by {\it Fermi}-LAT \citep[see the studies
  on Cen A, NGC 1275, and M87,][respectively]{2010ApJ...719.1433A,2009ApJ...699...31A,2009ApJ...707...55A}.
For comparison, typical bulk Doppler factors and magnetic fields
of BL Lac objects are $\delta =12$ and $B = 1.5\times10^{-2}$ G for
Mrk 501 \citep{2011ApJ...727..129A}, while for the extreme case of PKS 2155-304 during flaring states, detailed
SSC modeling gives $\delta \cong 100$ and $B \cong 0.1$ G
\citep{2008ApJ...686..181F}.\\
We note that the adopted bulk motion value implies a jet radiation cone $\theta_{rad}=1/\Gamma\sim24^\circ$. Thus,
the observer line of sight falls roughly inside the radiation cone of
the emitting region and NGC 6251 is aligned in the sense that it is within $1/\Gamma$.
Once again, we underline that the
definition of MAGN is strictly related to the expectation of
detecting radiation from a blazar-like (i.e. highly relativistic)
region even in those radio sources whose jet physical axis is oriented
at larger inclination angles ($\theta\gtrsim 10^\circ$) with respect to the line of sight. 
\\
The values of the main parameters are similar to the results of the SSC model
fitting of NGC~6251 in \citet{2003ApJ...597..166C}, where $\theta=18\degr$ and
$\Gamma=3.2$ were assumed.
The particle-to-magnetic field energy density ratio is $U'_e/U'_B
\ga 400$, where $U'_e=m_ec^2 n_e\langle\gamma\rangle$ ($m_e$ is the
electron mass, $\langle\gamma\rangle$ is the average electron Lorentz
factor) and $U'_B=B^2/8\pi$. The violation of the minimum energy
assumption is then rather severe already without considering any
contribution from the protons.\\
In summary, the SSC model provides a good overall fit to the data, but
requires a large departure from the equipartition between relativistic
electrons and magnetic field and a relatively slow speed of the
jet. We note incidentally that \citet{2000PASJ...52..989S} discussed a possible sub-parsec
scale acceleration of the jet from $\beta\sim 0.13$
$\approx0.5$ pc to $\beta\sim 0.42$ at $\approx 1$ pc. According to
this, we should be observing the jet before it becomes relativistic.\\
How robust are these results? The source orientation (with the related
quantities, $\Gamma$ and $\delta$) is indeed a crucial assumption.
A smaller $\theta$, say $\sim10\degr$, while
requiring a higher bulk Lorentz factor, $\Gamma\approx 15$, is
disfavored by radio observations: this $\theta$ would imply either a large linear size of the source,  $\ga
5.5$ Mpc, or a significantly bent jet \citep{2003ApJ...597..166C}.\\
We note that $\theta$ and $\Gamma$, as also $R$, affect in the same way the synchrotron and IC curves and will not change our results for the particle to magnetic field relative energy densities.
Beyond the uncertainties related to the assumed parameters (eg
$\theta$), we also considered those cause by observational
uncertainties and examined three possibilities: (i) the
$\gamma$-ray flux was overestimated bacause of contamination from an
external contribution \citep[for example the kpc jet or the lobes, as
  in the case of Cen A,
  see][]{2010Sci...328..725F,2010ApJ...719.1433A,2008ApJ...679L...9B};
(ii) both the $\gamma$-ray and X-ray fluxes are overestimated, because
of the limits of the X-ray and $\gamma$-ray observatories in resolving
the dimensions of the emitting region ($R\approx 10^{17}$ cm); and (iii)
the flux contamination introduces uncertainties in the values of
spectral indices . As a test, we considered the case where the fluxes
were overestimated by at most a factor of 5. 
We assumed an intrinsically harder source, assigning to 
$\alpha_1$ and $\alpha_2$ the values  $0.84$ and $1.4$ respectively,
based on the uncertainties
on the X-ray and $\gamma$-ray spectral indices.
Summarizing the results for the three cases, we have that $B$, $K$, and
$\gamma_b$ vary in the ranges $(3.7-6)\times 10^{-2}$ G,
$(1.5-0.25)\times10^{6}$ cm$^{-3}$, and $(0.8-2)\times 10^4$
respectively. It follows that $U'_e/U'_B$ may be reduced at most an
order of magnitude but still the jet remains particle-dominated.\\   

\begin{table}
\caption{NGC~6251 nuclear SED - models.}
\label{t3}
\begin{center}
\begin{tabular}{lllr}
\hline
\hline                    
\noalign{\smallskip}
Parameters  &Model 1   &\multicolumn{2}{c}{Model 2}\\
            &SSC       &Layer  &Spine\\
\noalign{\smallskip}
\hline
\noalign{\smallskip}
R (cm)                &$1.2\times10^{17}$    &$8.0\times10^{16}$  &$1.0\times10^{16}$\\
L (cm)                &--                    &$2.5\times10^{16}$  &$1.0\times10^{15}$\\
K (cm$^{-3}$)          &$1.5\times10^6$      &$7.0\times10^4$     &$7.1\times10^4$\\         
$\gamma_{min}$         &$250$                &$60$                &$100$\\
$\gamma_{max}$         &$1.4\times10^5$      &$4.1\times10^4$     &$2.0\times10^4$\\
$\gamma_b$            &$2.0\times10^4$      &$7.0\times10^3$     &$3.0\times10^3$\\
$p1$                  &$2.76$              &$2.76$               &$2.1$\\
$p2$                  &$4.04$              &$4.04$               &$3.0$\\
B (G)                 &$3.7\times10^{-2}$   &$0.7$
&$1.8$\\
$\Gamma$              &$2.4$               &$2.4$                &$15$\\ 
$\delta$              &$2.4$               &$2.4$                &$0.7$\\
\noalign{\smallskip}
\hline
\end{tabular}
\end{center}
\end{table}

\subsection{Structured jet models}

\citet{2000A&A...358..104C} pointed out that the luminosities of FRI radio sources exceed those predicted
by a simple unification scenario, which assumes that FRI  radio galaxies
represent the de-beamed counterparts of BL Lac objects whose
emission is produced, via SSC, by a single region.
Moreover, model fitting of NGC~6251, as also of other radio galaxies,
do require (modest) beaming factors
\citep{2001MNRAS.324L..33C,2009ApJ...699...31A,2009ApJ...707...55A,2010ApJ...719.1433A}.
If the condition of a homogeneous one-zone emitting blob is relaxed,
other possible interpretations are viable.  For example,
\citet{2005A&A...432..401G} proposed a stratified
jet structure, with a  fast narrow spine responsible for the BL Lac-like emission, and a slower layer, whose emission would dominate that of
the fast spine at larger angles, i.e. as in FRI radio sources.  
If true, there would be a
strong feedback  between the two components. The spine provides additional seed photons via synchrotron
emission that can be IC scattered by the relativistic electrons of the
layer (and vice versa).
Similar solutions, assuming a complex velocity axial/radial structure
in jets, were also proposed by other authors
\citep{2002ApJ...578..763S,2003ApJ...594L..27G}.
The presence of these structures are also supported by the transversely resolved brightness profile
observed in parsec scale jets \citep{2004ApJ...600..127G,2006ApJ...646..801G}.\\ 
In view of the new \fermi detection and additional (Spitzer and
Planck) data in the mm-IR band, we reconsider the spine-layer scenario
previously proposed for NGC~6251 in \citet{2005A&A...432..401G}.
The `fit' to NGC~6251 nuclear SED thus obtained is shown in Fig.\ref{f4} ({\it right panel}). The geometry of the two regions is cylindrical ($L$ is the cylinder length), with the spine nested inside the hollow layer cylinder. The spine parameters in Table \ref{t3} are chosen to reproduce a standard BL Lac object, with $\Gamma_{spine}=15$ for a region of radius $R_{spine}=10^{16}$ cm  in a magnetic field $B_{spine}=1.8$ G. The spine EED is a broken power-law with $p_1=2.1$ and $p_2=3.0$ and an energy break $\gamma_{b,spine}=3.0\times10^3$. We keep the same spectral shape as before for the EED of the layer ($p_1=2.76$ and $p_2=4.04$),
and the same $\delta$ and $\theta$. 
The red and black solid (dashed) curves in Fig. \ref{f4} ({\it right
  panel}) represent the layer (spine) synchrotron and SSC
contributions, respectively. The blue long-dashed curve is the layer IC
emission of the spine synchrotron photons. The interaction between the
spine and layer ensures the efficiency in reproducing the observed
X-ray to $\gamma$-ray emission and requires a low particle density
($K_{layer}=7\times10^4$ cm$^{-3}$), in favor of a larger magnetic field
($B_{layer}=0.7$ G). On the other hand,  the debeaming
($\delta_{spine}\sim0.7$) of the spine emission at $\theta=25\degr$
hides the presence of the emission from the fast flow.
  When the jet
axis inclination with respect to the line of sight is reduced, to say
$\theta\lesssim5^\circ$, the blazar-like emission from the spine dominates the SED
\citep[as also shown in Figure 5 in][for similar parameter values]{2005A&A...432..401G}.\\
A critical point is that the spine-layer model assumes a flux
variability in the spine radiation similar to that observed in
blazar sources. Rapid changes in the spine synchrotron flux should be
reflected by those of Comptonized emission coming from the layer, on
timescales of about
$t_{var}\sim(R_{spine}/c)\delta_{layer}\approx 0.02\ {\rm yrs}$. In
NGC~6251, no clear indication of variability in $\gamma$-rays has been
found, while the X-ray emission has varied on a shorter than two-year
timescale. 
The lack of significant variability in the $\gamma$-ray band, however,
cannot by itself rule out the model. Even without considering the source detectability,
clearly the volume of the layers could partly dilute the variability and
of course the real jet structure is likely to be more complex than
assumed in this simplified description.\\ 
Though beyond the aim of this work, we recall that more
recently two classes of models have been proposed to explain the high
energy emission in blazars and radio galaxies. In the first case, beams
of particles or'jets in a jet' emitting also off the jet axis are
invoked to explain the observed TeV flares in both blazars
\citep[eg][]{2009MNRAS.393L..16G,2009MNRAS.395L..29G} and the misaligned AGN M87
\citep{2010MNRAS.402.1649G}. In the second scenario, the high-energy
emission is produced by colliding shells with different bulk
motions. In a schematic and simplified view, (internal) shocks are
caused by the broad range of velocities characterizing the different "shells": the
slower ones would naturally lead to emission at larger viewing angles \citep{2010arXiv1011.0169D}.

\subsection{Jet power}
While from the modeling of the SED and its variability properties it is not possible to discriminate between the SSC and spine-layer scenarios, these have different implications as far as the jet energetics is concerned as inferred here.
Jet powers for electrons ($L_e$), protons ($L_p$), and Poynting flux
($L_B$) are estimated as,
\begin{equation}
L_{i}=\pi R^2 \Gamma^2 \beta c U'_i 
\end{equation}
where $U'_i$ is the energy density, in the comoving frame, of
electrons, $U'_e$, cold protons, $U'_p=m_p c^2 n_p$ ($m_p$ proton
mass), and magnetic field, $U'_B$. Here, we assume that there is one
cold proton per emitting electron ($n_p=n_e$).
The radiation luminosity is
\begin{equation}
L_r=\pi R^2 \Gamma^2 c U_{rad}'\sim L' \Gamma^2,
\end{equation}
where $L'$ is the intrinsic total (synchrotron and IC) luminosity. 
The kinetic, magnetic, and radiative powers for the two applied models,
SSC and spine-layer, are shown in Table \ref{t4}.\\

\begin{table}[h!]
\caption{NGC~6251 Jet Powers for the SSC and spine-layer models.}
\label{t4}
\begin{center}
\begin{tabular}{l l l r}
\hline
\hline
\noalign{\smallskip}
                          &Model 1             &\multicolumn{2}{c}{Model 2}\\
                          &SSC                 &Layer     &Spine\\
\noalign{\smallskip}
\hline
\noalign{\smallskip}
$L_e$ (erg s$^{-1}$)      &$1.6\times10^{44}$  &$3.9\times10^{42}$    &$4.8\times10^{43}$\\
$L_p$ (erg s$^{-1}$)      &$4.9\times10^{44}$  &$5.2\times10^{43}$    &$2.2\times10^{44}$\\
$L_B$ (erg s$^{-1}$)      &$3.6\times10^{41}$  &$2.3\times10^{43}$    &$4.8\times10^{43}$\\
$L_{kin}$ (erg s$^{-1}$)  &$6.5\times10^{44}$  &$5.6\times10^{43}$    &$2.7\times10^{44}$\\
$L_r$ (erg s$^{-1}$)      &$2.0\times10^{43}$  &$2.4\times10^{43}$
&$9.0\times10^{43}$\\
\noalign{\smallskip}
\hline
\end{tabular}
\end{center}
\end{table}

As noted before, the SSC model implies a strong particle dominance ($L_e\approx 1.6\times10^{44}$ erg s$^{-1}$ and $L_p\approx 5\times10^{44}$ erg s$^{-1}$) over the magnetic field ($L_B\approx 3.6\times10^{41}$ erg s$^{-1}$).
The radiative power is $L_r\approx 2\times10^{43}$ erg s$^{-1}$, about
an order of magnitude lower than the kinetic one. 
In this case, the bulk of the jet power is conserved and goes into the formation of the large radio structures.
The values are in the range of powers found by modeling the SEDs of
typical BL Lacs
\citep{2008MNRAS.385..283C,2010MNRAS.402..497G,2010MNRAS.401.1570T},
even though we note that NGC~6251 is in the low tail of $L_B$ values for BL Lacs \citep{2008MNRAS.385..283C}. \\ 
The SSC model - if correct - would have two interesting implications: (i)
$L_e$ and $L_p$ are comparable to those of BL Lacs with similar $L_r$, despite of the
different Lorentz factor, and thus the jet is relatively heavier; 
and (ii) the low Poynting flux seems to exclude a magnetically confined jet.
On kpc scales, the jet expands at variable lateral velocity, and exhibits the presence of re-confinement sites \citep{1984ApJS...54..291P}. Several authors \citep{1984ApJS...54..291P,1997A&A...324..870M,2005MNRAS.359..363E} have shown that the pressure exerted by the extended (out to 100 kpc) halo of X-ray emitting gas around the radio source can account for the jet confinement on kpc scales. On the other hand, in the inner few arcseconds the jet is rapidly expanding and thermal confinement would require an X-ray luminosity incompatible with the observed one. An initial regime of free expansion seems then more likely \citep{1984ApJS...54..291P}, requiring a re-collimation mechanism.\\
In contrast,  in the spine-layer model  the jet is faster and less
heavy. Thus, magnetic fields could play a fundamental role in its confinement. 
The total jet kinetic power derived from the spine-layer model
parameters is just slightly smaller ($L_{kin}=L_e+L_p\approx
3\times10^{44}$ erg s$^{-1}$) than $L_{kin}$ of the SSC model. However, in this scenario, because of
the higher energy density in seed photons than in the SSC
model, the required emitting particle density is lower. This leads
also to quasi-equipartition between $U'_e$ and $U'_B$ \citep{2005A&A...432..401G}.
The bulk of the jet kinetic power is carried by the fast spine. 
The radiative dissipation ($L_r=9\times10^{43}$ erg s$^{-1}$) is rather high in the spine. The jet converts about 30\% of its total power into radiation, and should then undergo a progressive deceleration, as also predicted by the Compton rocket effect \citep{2005A&A...432..401G}, while we recall that NGC~6251 has an extraordinary linear extension of $\sim1.9$ Mpc.\\

\subsection{Checking the  jet kinetic power }

The inferences about $L_{kin}$ rely on some key assumptions, the ratio
of the number density of
cold protons to that of electrons above all. A check on these estimates can be
done using an alternative approach, based on the measure of the $pdV$
work done by the source in forming the lobes in the hot surrounding
gas
\citep{1999MNRAS.309.1017W,2002MNRAS.331..369F,2006MNRAS.372...21A}.
Here, we adopted the correlation reported by \citet{1999MNRAS.309.1017W}
\begin{equation}
L_{kin}=3\times 10^{21} f^{3/2} L_{151}^{6/7} {\rm\ erg\ s^{-1}},
\end{equation}
where $L_{151}$ [erg s$^{-1}$ Hz$^{-1}$ sr$^{-1}$] is the
monochromatic radio power at 151 MHz. In the revised formula considered here, a factor $f$ takes into account possible systematic underestimates intrinsic to the technique. 
\citet{2007MNRAS.376.1849H} estimated $f$ to be in the range between 10 and
20 for a sample of FRI and FRII sources. Given the 151 MHz luminosity
of NGC~6251, $L_{151}=2.46\times10^{31}$ erg s$^{-1}$ Hz$^{-1}$
sr$^{-1}$, the kinetic power, setting $f$ respectively equal to 10 and
20, is in the range  $7.5\times10^{43}$--$2.2\times10^{44}$ erg
s$^{-1}$, just slightly below the value of $L_{kin}$ found for the two
models, thus in agreement with the above estimates.

\section {Jet power versus accretion}
The SED of NGC~6251 in the optical-UV band is dominated by the
non-thermal emission of the jet. No sign of a big blue bump is
identified and there is no strong evidence in the X-ray spectrum of
a broad Fe K$\alpha$ emission line, which are both considered to be signatures of
emission by an accretion disk. 
The nonthermal flux can be used to derive an upper limit to the disk
luminosity, $L_{disk}\lesssim10^{43}$ erg s$^{-1}$
\citep{1999ApJ...515..583F,2009ApJ...699..626H}, which
in units of
Eddington luminosity, $L_{Edd}$, is $L_{disk}/L_{Edd}\lesssim 10^{-4}$ for an estimated black hole mass $M_{BH}=6\times10^8\
M_{\sun}$ \citep{1999ApJ...515..583F}. The sub-Eddington regime could
indicate the presence of a radiatively inefficient disk \citep[here we generically refer to the class of models as radiatively inefficient accretion
  flows, RIAFs,
  see][for a review]{2002luml.conf..405N}. 
\citet{2001A&A...379L...1G} demonstrated that the FRI--FRII dividing line in the
radio luminosity vs optical galaxy luminosity plane \citep[proposed by][]{1996AJ....112....9L}  can be expressed in
terms of a critical accretion rate of $10^{-2}-10^{-3}$, in units of Eddington
accretion, and suggested that this might reflect a change in the accretion
mode from a standard optically thick geometrically thin efficient
Shakura-Sunyaev disk \citep{1973A&A....24..337S} to a RIAF.\\
\citet{1999ApJ...515..583F} estimated for NGC~6251 a Bondi accretion
rate of
 $\dot{M}_{Bondi}\sim 5 \times 10^{-4}$ M$_{\sun}$ yr$^{-1}$ and a corresponding accretion
power of $P_{accr,Bondi}=\eta \dot{M}_{Bondi}c^2\sim 3\times 10^{42}$ erg s$^{-1}$,
assuming an efficiency for the conversion of the accreted
rest mass into energy of $\eta=0.1$. Given the observed (non-thermal) luminosity,
they argued that a RIAF would require an accretion rate $10^2-10^3$ higher.
However, as they pointed out, this accretion rate value is likely to be a lower limit based on lower
limits to the pressure and density of the interstellar matter (ISM). 
If we assume that the jet is completely powered by the accretion
process, the jet kinetic power provides us with an even higher
lower-limit\footnote{Indeed, this lower limit relies on the assumption
  that there is a
proton component, which seems however supported by independent jet
power estimates (see Sect. 5.4). In addition, we assume that the proton
component is not due to the jet entrainment of intergalactic matter.}  to the accretion power
$P_{accr}\gtrsim L_{kin}\approx 10^{44}-10^{45}$ erg
s$^{-1}$. When we adopt typical gas densities found for the central regions of elliptical galaxies,
$n_d\approx 0.1-0.5$ cm$^{-3}$
\citep{2001ApJ...547..731D,2005MNRAS.364..169P}, we obtain a Bondi accretion rate
 $\dot{M}_{Bondi}\sim 5\times 10^{-2}$ M$_{\sun}$ yr$^{-1}$,
which gives $P_{accr,Bondi}\approx 10^{45}$ erg s$^{-1}$. 
If this is the
case, a radiatively inefficient disk can in principle provide the
required jet power, though the mechanism that
converts the accretion into the jet power is very efficient.
A study of nearby X-ray luminous elliptical galaxies have
 found that $P_{accr,Bondi}$ and the jet power, inferred
 from the work done to expand the cavities observed in the surrounding
 X-ray gas, have similar values. This suggests that a significant fraction of the mass
 entering the Bondi accretion radius eventually goes into the relativistic jet \citep{2006MNRAS.372...21A}.
Interestingly, it has been proposed for blazars \citep[][and
  references therein]{2008MNRAS.385..283C,2010MNRAS.402..497G} that the
jet power maintains a linear dependence on $\dot{M}/\dot{M}_{Edd}$
  with a transition to a quadratic dependence at low accretion rates, as also expected in RIAF solutions \citep{1995ApJ...452..710N,1997ApJ...477..585M}. 

\section{Summary}
We have presented a study of the broad-band nuclear SED of
the radio galaxy NGC~6251, one of the MAGNs detected by {\it
  Fermi}-LAT. In agreement with previous studies, the nuclear SED is
dominated by non-thermal emission related to the sub-pc jet. A nuclear
origin appears to be the most likely hypothesis for the $\gamma$-ray
emission. This is also supported by the results of our X-ray analysis
of archival XMM-Newton, Chandra, and Swift observations. Both models,
SSC and spine-layer, adopted to reproduce the observed SED, provide a
good overall fit. In the following, we summarize the main results of
the SED modeling.\\
\\
A single-zone SSC model assumes a slow and heavy jet. The particle
power dominates over the magnetic field power by about three orders of magnitude.
This seems to rule out a magnetically accelerated and confined
jet. The X-ray halo surrounding the jet can account for its
confinement on kpc scales but the core region remains poorly constrained. A low bulk Lorentz factor, when a one-zone SSC model is adopted,
is also shared with other MAGNs detected by {\it Fermi}-LAT
\citep[][]{2009ApJ...699...31A,2009ApJ...707...55A,2010ApJ...715..554K,2010ApJ...719.1433A}.
This suggests the presence of at least a second site of $\gamma$-ray
emission in addition to the one we observe when the jet is pointing
directly toward us \citep{2000A&A...358..104C}.
A significant reduction of the jet inclination angle, which would
allow higher bulk Lorentz factors, is unsupported by the radio
observations. These results should not be significantly
affected by an overestimate of the $\gamma$-ray flux of a factor of a few
($\approx 5$). \\
A structured jet with a fast inner component and a slower
  layer is a possible explanation. In the spine-layer model proposed
  in \citet{2005A&A...432..401G}, the jet is relatively light and could
  be magnetically confined. The fast inner component  carries the bulk of the jet power.
On the other hand,  the strong radiative dissipation
  ($L_r\sim0.3\,L_{kin}$) is at odds with the Mpc-length of the jet
  and the flux variability predicted by the model in the high energy
  band has not yet been observed.\\
The derived jet powers, $\approx 10^{44}$ -- $10^{45}$ erg s$^{-1}$, are model-dependent and rely on some important assumptions. However, similar values were derived using the 151 MHz luminosity \citep{1999MNRAS.309.1017W}.\\
The disk component appears to be completely hidden by
  the jet emission. The sub-Eddington luminosity regime ($L_{disk}/L_{Edd}\lesssim 10^{-4}$) could be
  related to the presence of a RIAF. Support to this hypothesis is provided by
  the estimates of the Bondi accretion rate, albeit with significant uncertainties
  in the ISM physical parameters. As has been found for blazar sources, the mechanism channeling
  the accretion power into the jet seems to work in a more efficient
  way ($L_{kin}/L_{disk}\gtrsim10$ and $L_{kin}\sim0.1-1 P_{accr}$).

\begin{acknowledgements}
We thank the anonymous referee for her/his contribution to improve the paper.
G. Migliori thanks A. Capetti and D. Thompson for useful comments, professor
A.J. Buras for hospitality at TUM-IAS and F. Civano for helpful suggestions.\\
This research has made use of the NASA/IPAC Extragalactic Database (NED) which is operated by the Jet Propulsion Laboratory, California Institute 
of Technology, under contract with the National Aeronautics and Space Administration. 
The authors would like to thanks the ASI Science Data Center  (ASDC) for providing on-line facilities for the Swift/XRT data analysis.
\end{acknowledgements}
\bibliography{mybib}{}

\newpage
\begin{table*}
\caption{NGC 6251 nuclear SED - data.}
\label{t1}
\begin{center}
\begin{tabular}{lcccl}
\hline
\hline
\noalign{\smallskip}
Frequency                &Flux                  &Flux error             &Reference     &Angular\\
(Hz)                     &(Jy)                  &(Jy)                     &              &Resolution\\ 
\noalign{\smallskip}
\hline
\noalign{\smallskip}
$3.26\times10^{8}$                   &$0.70$      &$0.10$                              &--    &{\small WENSS}$^{(a)}$ (55$\arcsec$)    \\                               
$4.09\times10^{8}$                   &$<0.30$     &--                                  &1        &3.7$\arcmin$\\
$6.09\times10^{8}$                   &$0.60$      &$0.10$                          &--       &{\small WENSS}$^{(a)}$ (56$\arcsec$)\\
$1.40\times10^{9}$                   &$0.50$      &$0.10$                            &--   &{\small NVSS}$^{(a)}$\\
$1.41\times10^{9}$                   &$0.36$      &--                                              &1                    &--\\
$1.67\times10^{9}$                   &$0.28$      &--                                &2          &{\small VLBI} ($<$3 mas)\\
$2.30\times10^{9}$                   &$0.25$      &--                                              &2                    &--\\
$2.30\times10^{9}$                   &$0.65$      &--                                  &3                 &{\small VLBI} (0.6 mas)\\
$2.70\times10^{9}$                   &$0.72$      &$0.02$                                    &1                    &3.7$\arcsec$\\ 
$4.75\times10^{9}$                   &$0.60$      &$0.03$     &--      &{\small VLA}$^{(a)}$\\
$5.00\times10^{9}$                   &$0.36$      &--                                              &2                    &--\\
$5.00\times10^{9}$                   &$0.13$      &--                 &4   &{\small VSOP} (0.5 mas)\\
$8.40\times10^{9}$                   &$0.72$       &$0.01$         &--        &{\small VLA}$^{(a)}$\\
$1.05\times10^{10}$                  &$0.55$       &$0.05$       &5  &--\\
$1.07\times10^{10}$                  &$0.90$       &--                                              &3                    &--      \\
$1.50\times10^{10}$                  &$0.34$       &--       &4       &{\small VLBA}  (0.5 mas)    \\
$1.50\times10^{10}$                 &$0.66$      &$0.01$         &--
&{\small VLA}$^{(a)}$\\   
$1.54\times10^{10}$                 &$0.85$                                  &$0.04$                                    &1                    &0.65$\arcsec$\\
$2.28\times10^{10}$                 &$1.30$
&$0.04$                                    &6                    &Wmap (0.88$\degr$)\\
$3.00\times10^{10}$                 &$1.34$           &$0.06$&7   &Planck ($32.65\arcmin$)\\
$3.30\times10^{10}$                 &$1.50$            &$0.07$                                    &6                    &0.66$\degr$\\
$4.00\times10^{10}$                  &$1.35$           &$0.13$
  &7     &Planck ($27\arcmin$)\\
$4.07\times10^{10}$                 &$1.50$                                  &$0.08$                                    &6                    &0.51$\degr$\\
$4.30\times10^{10}$                 &$0.50$       &$0.02$    &--    &{\small VLA}$^{(a)}$\\
$6.08\times10^{10}$                 &$1.20$                                  &$0.10$                                    &6                    &0.35$\degr$\\
$7.00\times10^{10}$                 &$0.84$            &$0.15$        &7 &Planck ($13.01\arcmin$)\\
$9.35\times10^{10}$                 &$0.70$            &$0.30$                                    &6                    &0.22$\degr$\\
$1.00\times10^{11}$                 &$0.82$            &$0.09$         &7 &Planck ($9.94\arcmin$)\\
$1.43\times10^{11}$                 &$0.70$            &$0.08$           &7   &Planck ($7.04\arcmin$)\\
$2.17\times10^{11}$                 &$0.47$            &$0.07$           &7   &Planck ($4.66\arcmin$)\\
$3.45\times10^{11}$                 &$0.42$            &$0.01$                                    &8                    &23$\arcsec$\\
$1.00\times10^{13}$                 &$4.97\times10^{-2}$          &--                             &9                    &4-11$\arcsec$ Total emiss\\
$1.30\times10^{13}$                 &$9.00\times10^{-2}$          &$2.00\times10^{-2}$            &10                   &0.5-2$\arcmin$\\
$2.00\times10^{13}$                 &$2.79\times10^{-2}$          &--                             &9                    &4-11$\arcsec$ Total emiss\\
$2.00\times10^{13}$                 &$0.81\times10^{-2}$          &--                             &9                    &4-11$\arcsec$ Sync. emiss.$^{(b)}$\\
$1.00\times10^{13}$                 &$1.64\times10^{-2}$          &--                             &9                    &4-11$\arcsec$ Sync. emiss.$^{(b)}$\\
%
$1.42\times10^{14}$                 &$1.08\times10^{-3}$          &$0.11\times10^{-3}$            &11                    &0.2$\arcsec^{(c)}$ \\
$1.85\times10^{14}$                 &$5.45\times10^{-4}$          &$0.54\times10^{-4}$            &11                    &0.2$\arcsec^{(c)}$  \\
$2.55\times10^{14}$                 &$3.14\times10^{-4}$          &$0.31\times10^{-4}$            &11                    &0.2$\arcsec^{(c)}$  \\
$3.69\times10^{14}$                 &$2.00\times10^{-4}$          &$0.20\times10^{-4}$            &11                    &0.2$\arcsec^{(c)}$  \\
$5.40\times10^{14}$                 &$1.34\times10^{-4}$          &$0.13\times10^{-4}$            &11                    &0.2$\arcsec^{(c)}$  \\
$5.45\times10^{14}$                 &$1.21\times10^{-4}$          &$0.12\times10^{-4}$            &11                    &0.2$\arcsec^{(c)}$  \\
$8.73\times10^{14}$                 &$3.30\times10^{-5}$          &$0.33\times10^{-5}$            &11                    &0.2$\arcsec^{(c)}$  \\
$1.24\times10^{15}$                 &$1.77\times10^{-5}$          &$0.18\times10^{-5}$            &11                    &0.2$\arcsec^{(c)}$  \\
$1.52\times10^{22}$                 &$<1.62\times10^{-10}$        &--                             &12                  &0.97$\degr\times$0.74$\degr$\\
$3.00\times10^{22}$                 &$4.10\times10^{-11}$         &$3.07\times10^{-11}$          &12                  &0.97$\degr \times$0.74$\degr$\\
$5.37\times10^{22}$                 &$1.85\times10^{-11}$         &$1.12\times10^{-11}$          &12                  &0.97$\degr \times$0.74$\degr$\\
$9.58\times10^{22}$                 &$1.83\times10^{-11}$         &$0.84\times10^{-11}$          &12                  &0.97$\degr \times$0.74$\degr$\\
$5.06\times10^{23}$                 &$1.23\times10^{-12}$         &--                             &12                  &0.97$\degr \times$0.74$\degr$\\
$1.66\times10^{24}$                 &$1.31\times10^{-12}$         &--                             &12                  &0.97$\degr \times$0.74$\degr$\\          
\noalign{\smallskip}
\hline
\end{tabular}
\tablefoot{
\tablefoottext{a}{ Archival data.} 
\tablefoottext{b}{ Extrapolated IR synchrotron flux \citep[][see text]{2009ApJ...701..891L} }. 
\tablefoottext{c}{ Optical-UV non de-reddened flux. }} 
\tablebib{ (1)~\citet{1977MNRAS.181..465W}; (2)
    \citet{1986ApJ...305..684J}; (3) \citet{1979ApJ...233L.101C}; (4)
    \citet{2000AJ....120..697S}; (5) \citet{1997A&A...324..870M}; (6)
    \citet{2009ApJS..180..283W}; (7) \citet{2011arXiv1101.2041P}; (8)
    \citet{2003AJ....126.2677Q}; (9) \citet{2009ApJ...701..891L}; (10)
    \citet{1990AJ.....99..476K}; (11) \citet{2003ApJ...597..166C};
    (12) \citet{2002ApJ...574..693M}.}
\end{center}
\end{table*}

\end{document}